\documentstyle[12pt]{article}
\begin{document}
\title{
Chaotic Traveling Waves in a Coupled Map Lattice\\
}
\author{
Kunihiko KANEKO
        \thanks{E-mail address : chaos@tansei.cc.u-tokyo.ac.jp } \\
        {\small \sl Department of Pure and Applied Sciences,}\\
        {\small College of Arts and Sciences}\\
        {\small \sl University of Tokyo, Komaba, Meguro-ku, Tokyo 153, JAPAN}
 \\
}
\date{}
\maketitle
\begin{abstract}
Traveling waves triggered by a phase slip in coupled map lattices are
studied.  A local phase slip affects globally the system, which is
in strong contrast with kink propagation.
Attractors with different velocities coexist, and
form quantized bands determined by the number of phase slips.
The mechanism and statistical and dynamical characters are studied with
the use of spatial asymmetry, basin volume ratio, Lyapunov spectra,
and mutual information. If the system size is not far from an integer multiple
of the selected wavelength, attractors are tori, while weak chaos remains
otherwise, which induces chaotic modulation of waves or a chaotic itinerancy of 
traveling states.  In the itinerancy, the residence time distribution obeys
the power law distribution, implying the existence of a long-ranged 
correlation. Supertransients before the formation of traveling waves are noted 
in the high nonlinearity regime.  In the weaker nonlinearity
regime corresponding to the frozen random pattern,
we have found fluctuation of domain sizes and Brownian-like 
motion of domains.  Propagation of chaotic domains by 
phase slips is also found.
Relevance of our discoveries to B\'{e}nard convection experiments 
and possible applications to information processing are briefly discussed.
\end{abstract}

\section{Introduction}

Spatiotemporal chaos (STC) is high-dimensional chaos which involves
spatial pattern dynamics.  It covers turbulent phenomena in general,
including B\'{e}nard convection, electric convection in liquid crystals,
boiling, combustion, magnetohydrodynamic turbulence in plasmas,
solid-state physics (Josephson junction arrays, spin wave turbulence,
charge density waves and so on), optics, chemical reactions with spatial 
structures, etc. It is also important in
biological information processing with nonlinear elements like neural dynamics.

The coupled map lattice (CML) is a dynamical system with discrete time (``map"),
discrete space (``lattice"), and a continuous state.
It usually consists of dynamical elements on a lattice
interacting (``coupled") among suitably chosen sets of other elements
[1-15]. The CML was originally proposed as a simple model for 
spatiotemporal chaos.

Modelling through a CML is carried out as follows:
Choose essential procedures which are essential for the spatially extended 
dynamics, and then 
replace each procedure by a parallel dynamics on a lattice.  The CML
dynamics is obtained by successive application of each procedure.
As an example, assume that you have a phenomenon, created by
a local chaotic process and diffusion. Examples can be seen
in convection, chemical turbulence, and so on. In CML approach, we reduce the 
phenomena into local chaos and diffusion processes.  If we choose a logistic 
map $x'_n (i) = f(x_n (i))$ ($f(y)=1-ay^2$) to represent chaos, and a discrete 
Laplacian operator for the diffusion, our CML is given by

\begin{equation}
x_{n+1}(i)=(1-\epsilon ) f(x_n (i)) + \epsilon /2 (f(x_n (i+1))+f(x_n (i-1)))
\end{equation}

One of the merits of the CML approach lies in its predictative power of novel
qualitative universality classes, without being bothered by the details of 
phenomenology. Classes discovered thus far include
spatial bifurcation, frozen random chaos, pattern selection with suppression 
of chaos, spatiotemporal intermittency, soliton turbulence, 
quasistationary supertransients, and so on [2-7].

In the present paper we report a novel universality class in CML,
which is related to recent experiments in fluid convection and liquid crystals:
(chaotic) traveling waves.
We study the qualitative and quantitative nature of the chaotic traveling
wave, with the help of the Lyapunov analysis and co-moving mutual information flow.

In our model (1), observed domain structures are temporally frozen
when the coupling $\epsilon $ is small ($<.45$), as has been
studied in detail in \cite{KK-PD}.
For larger couplings, domain structures
are no longer fixed in space, but can move with some velocity.
For weak nonlinearity ($a<a_{ps}\approx 1.55$), the motion of a 
domain is rather irregular and Brownian-like,
while pattern selection yielding regular waves is found
for larger values of the nonlinearity. 
These two regions correspond to the frozen random phase
and (frozen) pattern selection in the weaker coupling regime \cite{KK-PD},
respectively.  Our novel discovery here is that the pattern is no longer
frozen but can slowly move.

The organization of the paper is as follows.  In section 2, the coexistence
of traveling-wave attractors with different velocities is shown.
The quantization of selected velocities is noted, and the basin volume for
each attractor is investigated.  The mechanism of traveling is
attributed to the existence of phase slips, phase-advancing units, as
will be studied in detail in section 3.
The traveling wave suppresses
chaos almost completely, as is confirmed by the Lyapunov analysis
in section 4.  When the size is not close to an integer multiple
of the selected wavelength, weak chaos remains, which induce the
modulation of traveling wave or a chaotic itinerancy over different traveling
wave states due to chaotic frustration in the pattern. The long-term 
correlation of the itinerancy is studied in section 5. The flow of 
information in the traveling wave is characterized by co-moving mutual 
information flow in section 6.  Quasistationary supertransients 
before falling on a traveling wave attractor are studied in section 7.
Switching among attractors by a local input is studied in section 8,
where it is shown that a single input can induce a transition of 
an attractor's velocity and thus affect the entire lattice.
In a weak nonlinearity regime,
the motion of a domain is no longer regular.  
The Brownian-like motion of domains is
studied in section 9. If the local dynamics is not chaotic, but
periodic with the period $2^n$,
we can have traveling kinks in the strong coupling regime.
These kinks are localized in space and do not have a global influence,
in contrast with the phase slips, as will be shown in section 10.
Discussions and a summary are given in section 11 \cite{KK-PRL}.

\section{ Selection of discrete velocities}

In the CML (1), only a few patterns with some wavenumbers are selected
for large nonlinearity ($a>1.55$ for $\epsilon =.5$).
Examples of attractors are given in Fig.1.  Besides the non-traveling
pattern, there are moving patterns which form a traveling wave. We note that 
such traveling attractors are not observed in the weak coupling regime
($\epsilon <.4$).  The selected velocities of the attractors in the examples
are rather low, in the order of $10^{-3}$.  

As can be seen in Fig.1, attractors with different velocities of waves coexist.
In the simulation, the admissible velocities $v_p$ for the attractors lie
in narrow bands located at $\pm v_1, \pm v_2, \cdots ,\pm v_k$ 
( e.g., $.8v_k <v_p <1.2v_k $ ).
For example, $ v_1 = .95 \times 10^{-4}$, $v_2 =1.9\times 10^{-3},$ and 
$v_3 =2.9\times 10^{-3}$, $v_4 =3.9\times 10^{-3}$, for $a=1.72$, 
$\epsilon=.5$, and $N=100$. No attractors exist with $v_p \approx 0$ but
$v_p \neq 0$.  There is a clear gap between the velocities of the attractors
in each velocity band; 
No attractors are found with different velocity from these bands 
around $v_k$.  For all parameters, $v_k$ is approximately proportional to $k$.

\vspace{.2in}

---Fig.1 ---

\vspace{.2in}

One might argue that this discreteness of the velocity bands may be an artifact of 
our model, which is discrete in both space and time.
Since the speed is very slow ( i.e., the order of $10^{-3}$ site per step),
it is not easy to imagine a mechanism to which our original discreteness
(the order of 1 site per step ) is relevant.  To examine
possible effects of the spatial discreteness, we have also simulated a
CML with a much longer coupling range, following the method of \S 7.5 in \cite{KK-PD};
i.e., a repetition of the diffusion procedure in the CML of $I_D$ times per  
local nonlinear mapping procedure.  With the increase of $I_D$, the 
range of the diffusion is increased, making our attractor spatially much
 smoother, approaching a continuous space limit.  Our traveling attractors 
have then, much longer wavelengths, and have higher quantized speeds.
For example the speed band at $v_1$ is amplified roughly 4 times by choosing
$I_D =8$ (for $a=1.72,\epsilon =.5, N=200$). Thus the 
discreteness in space is not relevant to the discrete selection of velocities.

The wavelength of a pattern is almost independent of the velocity
of an attractor.  The velocity is governed not by the (spatial) frequency but
by the form of the wave.
Since our model has mirror symmetry, a traveling wave attractor 
must break the spatial symmetry. The wave form is spatially 
asymmetric. Here this spatial (a)symmetry is not a local property.
Indeed, the waveform differs by domains of unit wavelength.  
The asymmetry is defined only through the average over the total lattice.
We have measured the spatial asymmetry by 

\begin{equation}
s \equiv <\frac{1}{N}\sum_{j=1}^N (x_{n}(j+1)-x_n (j))^3>
\end{equation}

with the long time average $<...>$.  This third power is
chosen just because it is the simplest moment of an odd power, since the first power 
$\sum_{j=1}^N (x_{n}(j+1)-x_n (j))$ vanishes due to the periodic 
boundary conditions.  In the present paper the velocity of an attractor is 
estimated by virtue of the following algorithm:
Find the minimum $k$ such that $\sum_{j=1}^N (x_{n+2m}(j)-x_n(j-k))^2$ is
a minimum.
Up to some value of $2m$, the minimum is found for the lattice displacement 
$k=0$. If the attractor is moving, at a certain delay $2m$, the minimum
is not obtained for $k=0$, but for $k =\pm 1$.  With the help of this
delay the speed of the pattern is estimated as
$\pm 1/(2m)$.  
In Fig.2, we have measured the above $2m$ over time 160000 steps, after 
discarding 10000 initial transients, to obtain an accurate for
the average velocity.  

The relationship between $s$ and $v_p$ is shown in Fig.2, for several values of
the parameter $a$.  If $a \stackrel{\geq}{\approx} 1.74 \approx a_{tr}$ \footnotemark , the 
relationship is rather simple.
The velocity of an attractor turns out to be 
proportional to its asymmetry $s$, as is plotted in Fig.2c)d).  Here
we note that there is a gap of velocity between frozen attractors ($v_p =0$) 
and traveling attractors.
For an attractor with velocity $v=0$, $s$ is zero within numerical accuracy.
Thus spatial symmetry is attained through the attraction to the non-traveling
attractor, starting from an initial condition with spatial asymmetry.  Again, 
this spatial symmetry is not a local but a global property.  Indeed, for each 
domain over a single wavelength, its waveform is not generally
mirror symmetric.  The 
asymmetry in each wave form is cancelled through the summation over the entire 
lattice. This attainment of self-organized symmetry is possible under the 
existence of traveling attractors. Indeed, for a weaker coupling regime 
without a traveling attractor, all attractors have a fixed structure 
\cite{KK-PD}, but they are not generally spatially symmetric.  Spatial 
asymmetry in the initial conditions is not eliminated in this case.

\footnotetext{See section 7 for $a_{tr}$, where possible mechanism for
the change of $s$-$v$ relationship at $a_{tr}$ is discussed.}

For $a<1.74$, the relationship between $s$ and $v_p$ is more complicated.
Attractors with $v_p =0$ can have a small non-vanishing asymmetry. 
The self-organized symmetry is not complete.  The velocity gap between 
frozen attractors  and moving ones is not seen.
Furthermore the linear relationship between $v_p$ and $s$ does not hold,
although we can see a band structure of velocities.  
One of the reasons of this complication is 
coexistence of attractors with different periods ( or frequencies),
as will be studied in the next section.
\vspace{.2in}

---Fig. 2 a)b)c)d)---

\vspace{.2in}


For random initial conditions, the probability to hit
an attractor with the velocity 0 or $\pm v_1 $ is rather high.  
In Fig.3, we have measured the basin volume ratio for attractors of
different velocities.  A band structure of admissible velocities is
found.  In each band there are discrete sets of admissible
velocities.  We have confirmed that there are many attractors with different 
velocities within each band by running a long-time simulation. 

\vspace{.3in}

---Fig. 3  a)b)-----

\vspace{.2in}

As the velocity
of an attractor increases,  its basin volume shrinks
rather drastically (see Table I).  The basin volume for $v_1$ is often rather 
large. As is shown in Fig.3, basin volumes decrease (approximately) in a 
Gaussian form with the velocity of the band ( $ exp( -K^2 \times const.)$ for 
attractors in the band $v_p =Kv_1$).
This Gaussian decrease is generally observed for any parameter value, 
although the basin ratios for the fixed and $v_1$ attractors may
vary.

\vspace{.3in}
Table I:  Velocity, asymmetry, and basin volume of fixed and traveling 
attractors.  $a=1.73$, and $\epsilon =.5$.  500 attractors from random initial
conditions are chosen to estimate the basin volume ratio.

\begin{tabbing}
111111111111111111\=111111111\=1111111111\=111111111\=11111111111\=1111111111\=11111\kill

velocity\>0\>$\pm v_1=.95  $\>$\pm v_2 =1.95$\>$\pm v_3=2.9 $\>$\pm v_4=3.8$\>$\times 10^{-3}$\\

asymmetry $s$\>0\>2.5 \>5.3 \>8.3 \>12 \>$\times 10^{-5}$ \\

basin volume ratio\>34.6\%\> 23.7\% \> 7.2\% \> 1.6\% \> 0.2\%\>\\

\end{tabbing}

Dependences of the velocity $v_p$ on the parameter $a$ and size $N$ are 
given in Figs.4-5.  In these figures,
we have measured the velocity by taking the average over 160000 time steps, 
after discarding 800000 transients starting from several initial conditions.  
These averaged velocities are plotted for $1.6<a<1.85$ 
for $N=50 $ ( in Fig.4), while they are plotted over $ 10 <N< 250$ for 
$a=1.73$, in Fig.5.
We can see the selection of discrete velocities rather well.  Velocities
lie in a narrow band around $v_k$ .  

\vspace{.2in}
----Fig. 4 ----

----Fig. 5 ----
\vspace{.2in}

As is given in Fig.5, selected velocities slowly decrease with the system 
size. Our system has a selected wavelength $R$, and the fractional part of
$\frac{N}{R}$ is rather essential for the nature of traveling
wave \footnotemark.  Except for this additional dependence, the velocity 
decrease is roughly fitted by $1/\sqrt N$ up to $N=200$. We also note that
higher bands successively appear ($v_k$ with larger $k$), with the increase of 
the system size, although the basin volume for such higher
bands is rather small due to the Gaussian decay as shown in Fig.3.  

\footnotetext{See section 4 for a novel dynamical state which appears
when there is a mismatch between the size and the wavelength.}

As has been reported, no traveling state has been observed
for $\epsilon < \epsilon _c \approx .402$. We note that the velocity does 
not go to zero as $\epsilon $ approaches $\epsilon _c $ from above. For
$.402  < \epsilon <.45$ the velocity lies between $1.0\times 10^{-3}$ and 
$1.8\times10^{-3}$ without displaying any symptoms of a decrease. Rather,
the basin volume for the traveling attractor vanishes with 
$\epsilon \rightarrow \epsilon _c$, which is the reason why only 
non-traveling attractors are observed for
$\epsilon <\epsilon _c$.  The basin volume for traveling states ( i.e.,
all attractors with non-zero velocities) is shown in Fig. 6.

\vspace{.2in}

---- Fig. 6 ----


\section{Phase Slips: Local units for global traveling wave}

To understand the mechanism of this velocity selection, we note that 
$x_n(i)$ oscillates in time. For $\epsilon =.5$, the oscillation is
almost periodic and the period is very close to 4.  Then one can assign 
a phase of oscillation to a lattice site $i$ relative to $(x_n(i),x_n(i+1))$.
It is possible to assign a phase change $m'\pi/2$ ($m'=\pm 1$) between the 
lattice site $i$ and the lattice site $i+j$ in a neighboring domain,
according to the order of the period-4 like motion. 

\vspace{.2in}

----Fig.7 ----
\vspace{.2in}

When there is a phase gap of $2\pi$ between sites $i$ and $i+\ell $, it is 
numerically found that this interval unit $[i,i+\ell ]$ 
maintains the traveling wave. For example, in the attractor with
velocity $v_1$ in Fig.1,
the oscillation is close to period 4, with slow quasiperiodic modulation.
For periodic boundary
conditions, the total phase change should be $2k\pi$.  The velocity is
zero for an attractor with $k=0$.  If $k=1$, there
must be a sequence of 5 domains with phases $0,\pi/2, \pi, 3\pi/2, 2\pi$
for corresponding lattice sites $i$ (Fig.7). This unit is a phase slip
of $2\pi$.  A phase slip with a negative sign is defined by the 
mirror-symmetric pattern of a positive one. An attractor with
the velocity of the band $v_k$ has exactly $k$ (positive)
phase slips, in other words, $2k\pi$ phase change over the total lattice.
($k$ equals the number of positive phase slips subtracted by negative ones).
Among attractors with the same number of phase slips, there can be various 
configurations of domains. The velocity variance among attractors within the 
same 
band depends on this configuration.  Since a phase slip is localized
in space, one might think that the movement is a local phenomenon
like soliton propagation.  This is not the case.  In the present case, this 
phase slip must pull all the other regions to make them travel, changing the 
phases of oscillations of all lattice points.  Thus a local slip influences
globally all lattice points.  Our dynamics gives a
connection between local and global dynamics.  One clear 
manifestation of the global aspect is the additivity of velocity.  In our
system, the velocity of the wave is proportional to the number of phase slips. 
This proportionality gives a clear distinction between
our dynamics and soliton-type dynamics, where, of course, the velocity of a 
soliton does not increase with the number of solitons present.

The phase of oscillation can clearly be seen with the use of spatial return 
maps, 2-dimensional plots of $(x_n(i),x_n(i+1))$.  When there is a phase slip,
the spatial return map shows a curve as in Fig.8.  A point $(x_n(i),x_n(i+1))$
rotates clockwise 
with time when there is a phase slip, while 
the point does not rotate for a non-traveling attractor.
When there are two phase slips, the rotation speed is twice
in addition to a slight change of the curve.  We note that the motion is 
smooth without any remarkable change of
rotation speed even when the lattice site lies at the phase slip
region.  In Fig.8., we have plotted spatial return maps for
attractors with 1,2,3, and 4 phase slips. In the figure the system size is 
64 lattice sites while each phase slip requires 16 lattice sites.  Thus the 
attractor with 4 slips (see fig. 8d)
consists only of a sequence of 4 repeated phase slip patterns. For 
attractors with less than 4 slips, there can be variable configurations
of domains other than the phase slips.  Depending on the configuration,
the spatiotemporal return maps are different.

When the return map shows a closed curve, the attractor is on a 
projection of a 2-dimensional torus.  This is the case
in the state consisting only of phase slips (see fig.8d).
In general, the curve is
not closed and the return map forms surface rather than a curve 
(see fig. 8.a-c), suggesting a higher-dimensional attractor.
Indeed, in  Fig 8c), for example, we can see clearly another frequency 
modulation. As will be confirmed in \S 4, the attractor is
on a higher-dimensional torus.  Frequencies of quasiperiodic
modulation depend on the number of phase slips and the
configuration of the domains.

\vspace{.2in}
----Fig. 8 ----

\vspace{.2in}

Some of the numerical results in the previous section are explained by
the phase slip mechanism in this section.

The proportionality between the asymmetry and the velocity can partially
be explained by the fact that each (positive)
phase slip gives rise to a certain contribution to the
asymmetry $s$.
In Fig.2c)d), however, the proportionality holds even in a level within each
band where the number of slips is identical.  The asymmetry can depend on 
the configurations of domains, besides the number of slips. So far it is
not clear why the proportionality holds even for such small changes of 
the asymmetry by the configurations, when the nonlinearity is large.

The basin volume vs. the velocity:  Let us assume that by a random initial
condition, a phase change between two domains
($\pm \pi /2$) are randomly assigned.  ( We have to impose the constraint that
its sum should be a multiple of integers of $2\pi$, but this is
not important  for the following rough estimate).
Then the probability for the sum of the phase change obeys the
binomial distribution.  For large $N$, the probability to have $K$ phase 
slips is estimated as $exp(-(K/\sigma )^2)$ with $\sigma \propto \sqrt N$.
Thus the probability to have $K$ phase slips is expected to decay 
with a Gaussian form with $K$.  Thus the Gaussian form of the basin volume
( see Fig.3b) and the $1/\sqrt N$
dependence of the velocity (in Fig. 4) are explained.

\section{ Chaos and Quasiperiodicity in the traveling wave}

To examine the dynamics of our attractors, Lyapunov spectra are
measured numerically through the product of Jacobi matrices \cite{KK-Physica},
and are plotted in Fig. 9.
For most parameters ($1.65<a<2.0$) and sizes, the maximal exponent is 
zero, irrespective of the velocity of attractors.
Thus chaos is completely eliminated by pattern selection,
and the attractor is a torus.
As is expected from the spatial return maps,
the attractor can be a higher-dimensional torus with more than one
null exponent.  Between attractors with $v=0$ and $v=v_p$, there are only 
slight differences in Lyapunov spectra ( see Fig. 9). 

\vspace{.2in}
----Fig. 9 ----

\vspace{.2in}

In our model a (traveling) pattern is selected such that it eliminates chaos
(almost) completely.  
If chaos were not sufficiently eliminated, it would be impossible to sustain
a spatially periodic pattern during the course of propagation.  Such 
elimination of chaos is not possible for every wave pattern, since our dynamics
has topological chaos.  In our system a wavelength $R$ is selected. When
the size $N$ is not close to a multiple of
the selected wavelength $R$, there can remain some frustration in
any pattern configuration, and weak chaos can be observed.

In narrow parameter regimes, we have seen a chaotic traveling waves for some 
sizes. For example, very weak chaos is observed around $a \approx 1.70$,
if $N$ is large, as is shown in Fig. 10.  
The corresponding spatial return map (see Fig.11) consists of curves 
( corresponding to regular traveling) and scattered points (corresponding to 
chaotic modulation). It should be 
noted that the chaotic modulation propagates in the opposite direction as
the traveling wave.

Lyapunov spectra are given in 
Fig.12, where few positive exponents exist in the traveling wave attractor.  
The number of positive exponents is small ($1 \sim 3$) compared with the 
system size $N$. Chaos, localized in a domain, propagates as a modulation
of the wave, as is shown in Fig.10. 
We also note that the spectrum is almost flat near $\lambda \approx 0$. 
The propagating wave leads to a Goldstone mode giving rise to a null
exponent.  

\vspace{.2in}
----Fig.10 ----
%

\vspace{.2in}
----Fig.11  ----

\vspace{.2in}
----Fig.12  ----


\section{ Chaotic Itinerancy of Traveling Waves}

As is shown in the previous section, there remains some frustration when forming
a wave pattern if the ratio $N/R$ is far from an integer.
When the frustration due to this mismatch between the size and
wavelength is large, it leads to spontaneous switching among 
patterns (see Fig. 13).  This spontaneous switching arises from chaotic 
motion of each pattern, and may be regarded as a novel class of chaotic 
itinerancy \cite{CI}.  Global interaction is believed to be necessary to obtain a chaotic itinerancy \cite{CI}.  Although the interaction of our model is local,
the phase slip globally influences all the lattice points, and thus satisfies
the condition for chaotic itinerancy.

Only few remnants of curves (corresponding to the traveling structure)
can be seen in the spatial return map (see Fig.14),while
scattered parts are more dominant than in the chaotic traveling in the
previous section.  The direction of rotation also changes with time, 
through the scattered points.  Both amplitudes and phases of 
oscillations are modulated strongly here.  

\vspace{.2in}
----Fig.13  ----

\vspace{.2in}
----Fig.14  ----
\vspace{.2in}

For the spontaneous switching, we need some kind of modulation of the 
wave.  Indeed, each 
waveform starts to be rather irregular in space and time in advance to the 
switching. The wavelength, on the other hand, is not affected by the course of 
this switching process. In general, there can be three types of modulation 
of the wave; frequency, phase, and amplitude modulations.  In our example, 
frequency is hardly modulated (as is seen in the invariance of wavelength 
through the switching),
while the phase modulation (following the amplitude one) is essential to 
the spontaneous switch of traveling states.

The switching occurs through the creation or destruction of a phase slip.
Frustration in a pattern leads to the distortion of a phase slip,
inducing chaotic motion.  This chaotic motion breaks the phase slip.
On the other hand, there can be the creation of a slip by chaotic
modulation of the phase of oscillation.
This creation or destruction of a phase slip is a local process,
but influences globally the velocity of the traveling wave.

In chaotic itinerancy, long time residence at a quasi-stable state is
often noted.  We have measured the residence time distribution of
a state with a given velocity.  As is shown in Fig.15, all the
residence time distributions $P_k (t)$ of a $k$-phase-slip state (for $k=0, 
\pm 1, \pm2$; i.e., $v_p =0,v_p =\pm v_1, v_p =\pm 2 v_1$) obey the power law
$P_k (t) \approx t^{-\alpha }$ with $\alpha \approx 1$. This power-law 
dependence clearly indicates the long time residence at each traveling state.
Similar power-law dependence of a quasistable state has already been found
for spatiotemporal intermittency in a CML \cite{KK-PD}, although the power 
itself is clearly distinct.

Lyapunov spectra for this frustration-induced chaos are shown in Fig.16. The
number of positive exponents is again very few (3 in the figure), whose 
magnitudes are very small.  The chaos by the frustration is very weak and 
low-dimensional.
The spectra have a plateau at the null exponent, implying the existence of 
a Goldstone mode by traveling wave.  As seen in the previous
section the  accumulation at null exponent is 
characteristic of a (chaotic) system with a traveling wave.

For larger system sizes, chaotic itinerancy of waves is hardly observed.
The system settles down to a frozen or traveling pattern after
transients.  Since the number of chaotic modes is few ($O(1)$), the
frustration per degree of freedom is thought to decrease with $N$.
The distortion due to the mismatch of phases is still there,
but it is distributed over a large size and is too weak to switch the
pattern. The remnant frustration in a traveling wave leads to chaotic 
modulation of wave as is studied in \S 4.

\vspace{.2in}

----Fig. 15 ----

----Fig. 16 ----

\vspace{.2in}

\section{ Co-Moving Mutual Information Flow}

Co-moving mutual information is often useful for measuring correlations in
space and time \cite{KK-Physica}.  From the joint probability 
$P(x_n(i),x_{n+m}(i+j))$, we have calculated the mutual information

\begin{math}
I(m,j)=\int dx_n (i)dx_{n+m}(i+j) P(x_n (i),x_{n+m}(i+j)) 
\end{math}

\begin{equation}
log \frac{P(x_n(i),x_{n+m}(i+j))}{P(x_n(i))P(x_{n+m}(i+j))}.
\end{equation}

In a traveling wave, we have peaks in $I(m,j)$ at $I(t,v_p t)$ for
an attractor traveling with $v_p $.  For a quasiperiodic attractor
the peak height does not decay with the time delay $t$, while it slowly decays 
for a chaotic attractor. The transmission of correlations can clearly be seen.

In a chaotic attractor, however, there is also propagation of small modulations
on the traveling wave.  As can be seen in Fig.10, this propagation is
in the opposite direction to the wave.  From the above mutual information,
this reverse propagation could not easily be measured so far.  The propagation 
of chaotic modulation implies the flow of information created by 
chaos \cite{Shaw}.  One way to measure this information 
creation may be the use of
three point mutual information with the use of 
$P(x_n(i),x_{n+m}(i+j),x_{n+m+\ell}(i+j)) $\cite{Ikeda}, while
another possible way of charactering a chaotic traveling wave is the use of 
the co-moving Lyapunov exponent \cite{KK-Physica}.  A slight increase of the
exponent at the traveling velocity 
is observed.  In our case, however,
chaos is too weak to give a quantitative distinction.

\vspace{.2in}

----Fig. 17 ----

\vspace{.2in}

The mutual information in the chaotic itinerancy decays with time and space,
without any peaks at some velocity.  By the switching process, all local
traveling structures are smeared out, leading to the destruction of
peaks in the mutual information at some velocities (see Fig.18).

----Fig. 18 ---


\section{ Chaotic Transients before the formation of Traveling Waves}

To fall on a traveling ( or fixed) attractor, the velocities of all
local domains of unit wavelength must coincide.  Thus it is expected that 
the transient time before falling on an attractor may increase with the system 
size. As for the transient behavior, our transient wave phase
splits into the following two regimes.

(i) For medium nonlinearity regime ($a<a_{tr} \approx 1.74$), the transient 
length increases at most with the 
power of $N$.  Indeed, local traveling wave patterns are formed within a few 
time steps.  Before hitting the final attractor, these local waves are 
slightly modulated
to form a global consistency.  The formation of a global wave structure 
occurs for time steps smaller than $O(N)$.  We need time steps in the order of 
$O(N)$ for the slight modulation to adjust the phases of all domains.
( see Fig. 19a)b) for spacetime diagram).

(ii)For larger nonlinearity ($a>a_{tr}$), there are long-lived chaotic 
transients before our system falls on a traveling-wave attractor. The 
transient length increases with the system size rather rapidly: the increase
is roughly estimated by $exp(const.\times N)$ \cite{Fred}, although some 
(number-theoretically)
irregular variation remains.  In the transient process, the dynamics
is strongly chaotic, and is attributed to "fully developed spatiotemporal
chaos" in \cite{KK-PD} \footnotemark.  Lyapunov spectra during the transients
are shown in Fig.20, in contrast with the spectra of an attractor. This 
transient process is quasistationary (see Fig. 19c for spacetime diagram);
No gradual decay is observed for dynamical quantifiers such as the
short-time Lyapunov exponent \cite{KK-THERMO} or Kolmogorov-Sinai entropy.
Such dynamical quantifiers fluctuate around some positive value, till
a sudden decrease occurs at the attraction to the regular attractor.
These observations are consistent with the type-II supertransients
often observed in spatially extended systems \cite{TRS}.
In a strong coupling regime, we have found traveling wave states
up to the maximal nonlinearity $a=2$.  Thus the fully developed
spatiotemporal chaos in this regime may belong to supertransients \cite{Fred}.

\footnotetext{If $a$ is not so large ( near $a \approx a_{tr}$), we have 
often observed some local traveling wave patterns during the transients.  This 
dynamics can be attributed to the
spatiotemporal intermittency of type-II \cite{INT}.}

We note that the linear relationship between the asymmetry $s$ and the velocity $v$
(in section 2) is seen only for $a>a_{tr}$.  This relationship
may be partially explained from the results in the present section,
although further studies are necessary for a complete explanation:
For $a<a_{tr}$, the pattern selection can occur locally, and some local distortion in 
the wave pattern may not be removed.  Then
spatial asymmetry can remain even for a non-traveling attractor, and
the $s$-$v$ relationship can be very complicated.
For $a>a_{tr}$, on the other hand, slight distortion in wave pattern leads to global 
chaotic transients.  Only patterns without distortion are admissible as attractors.  
The $s$-$v$ relationship may be expected to be monotonic and simpler. 

\vspace{.2in}

----Fig. 19  ----

-----Fig. 20 ----

\vspace{.2in}

\section{ Switching among attractors with different velocities}

By a suitable input at a site at one time step, we can make an external 
switch from
one attractor to another (with a different velocity). By a local input, the 
structure of an attractor is changed over the whole lattice. Local 
information by an input is transformed into a {\sl global} wave pattern 
(Fig.21). In the medium nonlinearity regime($a<a_{tr}$), the switching process 
occurs within a short time, without any global chaotic transients.

As is expected this switching is easily attained by applying an input
at site(s) in a phase slip.  For example, assume that the phases at neighboring
5 domains are given by $[0,+\pi /2, +\pi , +3\pi /2, 2\pi ]$.  By applying an
input at site(s) of the third domain, the phases can be switched to
$[0,+\pi /2, 0, -\pi /2(=3\pi /2), 0]$.  Thus a phase slip is removed,
leading to a switch to an attractor with $v\rightarrow v_{next}=v-v_1$.
We can control a switch by choosing an input site and value so that
the number of phase slips is in(/de)creased.  

For larger nonlinearity ($a>a_{tr}$), the chaotic
transient lasts for many time steps during the course of switching.
In this case the control of switching is almost impossible; it is hard to 
predict the length of the switching process or the attractor after the switch.
This type of chaotic transients in the search for an  attractor can be
seen in some models with chaotic itinerancy \cite{CI} and in the neural 
activity in an olfactory bulb \cite{Freeman}.

\vspace{.2in}

-----Fig. 21 -----

\section{Fluctuating Domain by Chaos}

In the weak coupling case, a frozen random pattern is observed \cite{KK-PD}
for weak nonlinearity ($a<1.55$).  In our strong coupling case, which phase
corresponds to it?  In this coupling regime, domains with variable sizes are 
again formed. These domains, however, are not fixed in space.  
The boundary of domains here  fluctuates in time.
Over some time steps some region moves in one direction locally, but then
it changes the direction of traveling  ( see Fig.22a).
In spatial return maps, the motion of $(x_n(i),x_n(i+1))$
along a curve  changes its direction with time.
The boundary motion is diffusive and Brownian-like (see Fig.22).  
Furthermore, the size of domains can also vary (chaotically).
Domain distribution is rather random.  We have plotted the spatial power 
spectrum $S(k)=<|\sum_j x_n(j)exp(ikj)|^2>$ with the temporal average $<...>$.
In contrast with the peaks corresponding to the selected wavelength
in the regular traveling wave regime, there are no clear peaks in the
spectra (see Fig 24).  The decay of the power spectra with the wavenumber is 
consistent with the diffusive motion of domains, while the decay in
the frozen random phase in the weak coupling regime is much slower,
due to the absence of such diffusive motion.

In this phase, some attractors have phase slips. Again a phase slip is defined 
by a unit with a sequence of domains of $2\pi $ phase advance.  In an
attractor with phase slips, the pattern moves (in average) in one direction 
with some fluctuation.  Generally the pattern has some average velocity
depending on the number of phase slips, although a fluctuating boundary of
domains brings about the fluctuation of the velocity.
Here chaos is not eliminated in large domains. In this case chaos is 
transported along with the traveling wave.  Chaos localized
in large domains moves together with the wave and in the same direction.
An example of a pattern is given in Fig. 23 with the corresponding
spatial return maps .

If $a$ is smaller, domains of various sizes coexist, while 
the appearance of larger domains is less frequent as $a$ approaches
$\approx 1.55$, where pattern selection sets in.

Chaos in the internal dynamics in a large domain is
confirmed by the Lyapunov spectra given in Fig. 25.  The slope of the spectra 
is small at $\lambda \approx 0$.  These exponents near 0 are thought to come 
from the diffusive motion of the domains.  Co-moving mutual information decays
exponentially in space and time (Fig.26), implying that there are
no remaining patterns in space and time.

\vspace{.2in}
----Fig. 22 ----

----Fig. 23 ----

----Fig. 24 ----

----Fig.25 ----

----Fig. 26 ----


\section{Propagating Kinks in Period-doubling Media}

To clarify the difference between the phase slips and conventional
solitons, we have studied our CML in the period-doubling regime with
a strong coupling.  As is known our CML exhibits the period-doubling
of kink patterns \cite{KK-PTP}.  In the lower coupling regime 
($\epsilon <.4$), these kinks are pinned at
their positions.  In the strong coupling regime, some kinks
can move when they form a phase gradient in one direction ( see for 
example fig.27).
This phase in-(de-)crease is possible if the period of the kinks is
larger than 2 \cite{kink}.  If the period is 4, for example,
there can be a series of domains with the increase of the phase
 $0, \pi /2, 3\pi /2, 2\pi$, separated by 3 kinks.  These kink patterns form 
a phase gradient, which drives them to move with a constant speed.

As for this phase advance, these moving kinks are similar to our phase slips.
However, the kinks here are completely local.  When there are
two kinks at a distant position, they move almost independently
with their original speeds, (until they collide).  See Fig.27, where elimination of
one kink by external input does not cause any change to the propagation
of the other kink.  Furthermore, there is no
discrete selection of speeds.  The speed of a kink gradually varies
with the phase gradient within the kink pattern. In Fig. 27, change of a tail
length at a kink leads to a slight change of speed.
Kinks here belong to the same class as those studied
in some partial differential equations like a $\phi ^4$ system. 
In an oscillatory medium (without chaos), we can expect
the existence of kinks with period-doubling as in the present example.

\vspace{.2in}
----Fig. 27  ----

\section{Summary and discussions}

In the present paper we have reported a traveling wave triggered
by phase slips.  The velocity of traveling attractors forms quantized bands
determined by the number of phase slips.  Frozen attractors (without
any phase slips) and traveling attractors with different velocities
coexist.  The velocity of each band increases linearly with the number of 
slips. When the nonlinearity is large, the proportionality between the
asymmetry
of a pattern and the velocity holds even within each band.
In this case,  (approximate) symmetry is self-organized for frozen attractors.

Through pattern selection
of domains with some wavenumbers, chaos is completely eliminated
leading to quasiperiodicity.  When there is a mismatch between the
size and the wavelength, remaining frustration leads to a chaotic motion of 
the wave.  If the frustration is large, chaotic itinerancy
over many traveling (and frozen) states is observed.  Our system itinerates 
over states with different velocities of traveling.
The residence time in each state
obeys a power law distribution.  If the frustration is not so large,
a chaotic traveling wave is observed, where the chaotic modulation is 
transmitted 
in the opposite direction to traveling wave.

It should be noted that a local phase slip affects globally the motion
of the total system.  This is in strong contrast with the kink
type propagation (also observed in our system in a non-chaotic region),
where it propagates as a local quantity.  The additivity of velocity of 
the wave with the number of phase slips is a clear
manifestation of the global nature.

By local external inputs, one can create or destroy a phase slip,
and to switch to an attractor with a different velocity.  
By the traveling wave, information is transmitted to the whole space
within the steps in the order of $O(N)$.  Thus the transformation from
local to global information is possible through this switching,
which may be useful for information processing and control.

We have noted two types of transient processes in the course of attraction 
to traveling states.  When the parameter for the nonlinearity is large,
a supertransient (with quasistationary measure) is observed
whose transient length increases exponentially with the system size, 
while such rapid increase is not seen in the medium nonlinearity regime
($a<1.75$).

In the weaker nonlinearity regime ($a<1.55$) corresponding to the frozen 
random pattern,
we have found fluctuation of domain sizes and Brownian-like 
motion of domains.  Coexistence of fluctuating domains and
phase slips is also noted.  In this random pattern, chaos is localized in
large domains, and propagates along the traveling wave.

We have analyzed the dynamics of the traveling wave with the use
of spatial return maps, Lyapunov spectra, and co-moving mutual information
flow.  When chaos is suppressed in the pattern selection regime, the 
maximal Lyapunov exponent is zero implying that the traveling 
wave attractor is on a torus whose dimension depends on the number of
phase slips.  This null Lyapunov exponent remains even in a chaotic 
wave or chaotic itinerancy.  This exponent is due to the Goldstone-type mode 
corresponding to
the traveling structure.  Through the mutual information flow, we note the 
creation and transmission of information by a chaotic traveling wave.
The chaotic modulation added on the traveling wave leads to the
possibility of information transmission, created by chaos \cite{Shaw}.

There is no apriori reason to deny the possibility of the traveling wave
in partial differential equation systems.
The existence of admissible velocity bands is thought to be due to
the suppression of chaos.  For convenience of an illustration, consider a
partial differential equation with two components 
\begin{equation}
\partial \vec{\phi } (r,t) / \partial t=\vec{F} (\vec{\phi }(r,t))+D \nabla ^2 \vec{\phi }(r,t).
\end{equation}
If there exists a traveling wave solution $\vec{\phi }(r,t)=\vec{f}(r-vt)$, 
it must satisfy
\begin{equation}
-v\vec{f'}=\vec{F}(\vec{f})+D\vec{f''}.  
\end{equation}
If this coupled second order differential 
equation has a periodic solution for a range of velocities $v$,
then the traveling wave $\vec{f}(r-vt)$ can be a solution of eq.(4). 
Generally speaking, nonlinear eq.(5) has a chaotic solution for some range
of $v$, and has windows of limit cycles 
in the parameter space ($v$) for chaos. 
This scenario implies the existence of admissible velocity bands for stable 
traveling wave solutions, as in our CML example. 

Traveling waves have often been studied in various experiments \cite{Kolo}.
Quite recently, Croquette's group has observed traveling waves in B\'{e}nard 
convection with periodic boundary condition.  Indeed this traveling wave 
is triggered by a unit of a $2\pi$ phase advance \cite{Croquette}.  They have 
also observed chaotic itinerancy
over different traveling states, when the motion in a B\'{e}nard cell is strongly
chaotic \cite{Yana}.  It is expected that this discovery belongs to the same 
class as our traveling wave.  It is interesting to check if attractors with 
different velocities coexist by applying perturbations to such experimental
systems.  A search for our velocity bands in experiments will also
be of interest.
Detailed comparison with our model and experiments will be important in future.

In the weak nonlinearity regime, the suppression of chaos
is not possible, where domains show chaotic Brownian motion without a 
traveling velocity. This type of floating domains has some correspondence 
with the dispersive chaos found in B\'{e}nard convection by Kolodner's group
\cite{Kolo2}. 

\vspace{.2in}

{\sl acknowledgements}

I would like to thank K. Nemoto, Frederick Willeboordse, 
K.Ikeda, M. Sano, S. Adachi, T. Konishi, J. Suzuki, for illuminating
discussions. This work is partially supported by Grant-in-Aids
(No. 04231102) for Scientific Research from the Ministry of Education,
Science and Culture of Japan.
A preliminary version of the paper was completed in February of 1992,
although I have had many fruitful discussions since then.
I would like to thank again  Frederick Willeboordse for discussions,
critical comments on the manuscript,  and
for sending me his Doctoral thesis in advance.
My stay at Paris in June 1992 gave me an exciting chance to 
encounter a beautiful experiment by Croquette's group.  I am
grateful to Hugues Chat\'{e} for his hospitality during my stay and discussions. 
This travel was supported by Grant-in-Aids (No. 04044020) for Scientific Research 
from the Ministry of Education, Science and Culture of Japan.

\pagebreak

Fig.1

Amplitude-space plot of $x_n(i)$ with a shift of time steps.  200 sequential 
patterns $x_n(i)$ are displayed with time (per 64 time steps),
after discarding 25600 initial transients, starting from a random initial 
condition.  $a=1.71, \epsilon =.5$, and $N=64$.
Four examples of attractors.
(a) $v_p =0$ (b)$v_p =-v_1$ (c)$v_p =v_1$ (d)$v_p =v_2$

\vspace*{.2in}
Fig.2  Asymmetry $s$ versus velocity $v_p$ for attractors 
started from randomly chosen 300 initial conditions.
The asymmetry and velocity are computed from the average of 
160000 steps after discarding 100000 initial transients.
$N=64$ and $\epsilon =.5$.
$(a) a=1.64 (b) a=1.69 (c)a=1.75 (d) a=1.8 $

\vspace*{.2in}
Fig.3  The basin volume for each attractor with the velocity $v_p$,
calculated from 2000 random samples, for $a=1.72, \epsilon=.5$, and $N=100$.
Obtained from 200000 steps after discarding 200000 steps.
The number of initial conditions fallen on the attractor with
the velocity $v_p$ is plotted as a function of $v_p$.
(b) The basin volume ratio for attractors in each velocity
band calculated from the data for (a).

\vspace*{.2in}
Fig.4  Velocities of attractors versus asymmetry $s$,
obtained with the algorithm in the text,
applied per 32 steps, over 32x5000 steps, after discarding
50000 initial transients.  Velocities from randomly
chosen 500 initial conditions are overlayed. $N=100$.  Data for
$a=1.66,1.67 \cdots,1.85$, are overlayed, for $\epsilon=.5$.
( additional data are included from fig.5 in \cite{KK-PRL}).

\vspace*{.2in}
Fig.5  The absolute values of velocities $|v_p|$ of attractors, 
plotted as a function of size $N$.
The velocities are computed with the algorithm in the text, 
applied per 32 steps, over 32x5000 steps, after discarding
50000 initial transients.  Velocities from randomly
chosen 50 initial conditions are overlayed. $a=1.73$, and $\epsilon=.5$.

\vspace*{.2in}
Fig.6  Basin ratio for traveling wave as a function of $\epsilon$,
for $a=1.69$.  Velocity of attractors from randomly
chosen 50 initial conditions are examined, to count the
number of attractors with $v \neq 0$.

\vspace*{.2in}
Fig.7  Space-Amplitude plot of $x_n (i)$. $a=1.72, \epsilon =.5$, and $N=64$.
100 steps are overlayed after discarding 10000 initial transients.
This wave pattern is traveling to the left direction.
(b) The same plot, shown per 4 steps.  Arrows indicate the phase of 
oscillation of the corresponding domains.

\vspace*{.2in}

Fig.8: Spatial Return Map:  $\{ x_n(1),x_n(2) \}$ are plotted over
the time steps $n=10001,10002, \cdots 210000$.
$a=1.70,\epsilon=.5$, and $N=64$.

\vspace*{.2in}
Fig.9: Lyapunov spectra of our model with $\epsilon = .5$, starting with
random initial conditions, discarding 50000 initial transients.  The 
calculation is carried out through the products of Jacobi matrices over
16384 time steps. N=64.
$ a=1.72$; for attractors with 2 phase slips (solid line) and one
phase slip (two examples; dotted lines), and
frozen attractors without a phase slip (two examples: broken lines)
 
\vspace*{.2in}
Fig.10:
Amplitude-space plot of $x_n(i)$ with a shift of time steps.  200 sequential 
patterns $x_n(i)$ are displayed with time (per 128 time steps),
after discarding 10240 initial transients, starting from a random initial 
condition.  $a=1.69, \epsilon =.5$, and $N=92$.

\vspace*{.2in}

Fig.11: Spatial Return Map:  $\{ x_n(1),x_n(2) \} $ are plotted over
the time steps $n=10001,10002, \cdots 210000$.
 $a=1.69, \epsilon =.5$, and $N=100$.

\vspace*{.2in}
Fig.12:
Lyapunov spectra of our model with $\epsilon = .5$, starting with
a random initial condition, discarding 50000 initial transients.  The 
calculation is carried out through the products of Jacobi matrices over
32768 time steps. N=100. a=1.69: for attractors with $v=v_1$ (solid 
or dotted line) and $v=0$ (broken line).

\vspace*{.2in}

Fig.13:
Amplitude-space plot of $x_n(i)$ with a shift of time steps.  200 sequential 
patterns $x_n(i)$ are displayed with time (per 1024 time steps),
after discarding 1024000 initial transients, starting from a random initial 
condition.  $a=1.69, \epsilon =.5$, and $N=51$.  


\vspace*{.2in}
Fig.14: Spatial Return Map:  $\{ x_n(1),x_n(2) \} $ are plotted over
the time steps $n=10001,10002, \cdots 210000$.
$a=1.69, \epsilon =.5$, and $N=51$.

\vspace*{.2in}
Fig.15 Residence time distribution for a state with $v \approx v_k$
in the chaotic itinerancy of traveling wave. The distribution is
taken over 819200 time steps after 20000 initial transients,
and sampled over 500 initial conditions.
 $a=1.69, \epsilon =.5$, and $N=51$.
(a) $k=0$ ( staying at a frozen state) (b) $k=1$ ( residence at a
one-phase-slip-state) (c) $k=-1$ ( residence at a
one-negative-phase-slip-state) (d) $k=2$ ( residence at a
two-phase-slip-state) 

\vspace*{.2in}
Fig.16:
Lyapunov spectra of our model with $\epsilon = .5$, starting with
a random initial condition, discarding 50000 initial transients.  The 
calculation is carried out through the products of Jacobi matrices over
32768 time steps. N=51. a=1.69: Spectra from three
different initial conditions are overlayed.
 
\vspace*{.2in}
Fig. 17: Co-moving mutual information flow for the logistic lattice (1),
obtained with the algorithm in the text. $I(m,t_c \times j)$ is plotted 
for $-10 \leq m \leq 10$ and $0 \leq j \leq 15 $ with coarsegraining time
$t_c=256$. 
The probability is calculated using 64 bins and
sampled over 5000x$t_c$ steps over the whole lattice.
(a) $a=1.69, \epsilon=.5$ and $N=100$: for a chaotic attractor with 
$v_p =v_1$.
(b) $a=1.69, \epsilon=.5$ and $N=100$: for an attractor with $v_p =-3v_1$.
(c) $a=1.69, \epsilon=.5$ and $N=100$: for an attractor with $v_p =0$.
\vspace*{.2in}

Fig. 18: Co-moving mutual information flow for the logistic lattice (1),
obtained with the algorithm in the text. $I(m,t_c \times j)$ is plotted 
for $-10 \leq m \leq 10$ and $0 \leq j \leq 16 $ with coarsegraining time
$t_c=256$. 
The probability is calculated using 64 bins and
sampled over 5000x$t_c$ steps over the whole lattice.
$a=1.69, \epsilon=.5$ and $N=51$
\vspace*{.2in}

\vspace*{.2in}
Fig.19: Space-time diagram for the coupled logistic lattice (1), with
$ \epsilon =0.5$, and starting with a random initial
condition.  If $x_{n}(i)$ is
larger than $x* $(unstable fixed point of the logistic map) ,
the corresponding space-time pixel is painted as
black (if $x>x_2 ^*\equiv (\sqrt{1+7a}-1)/(2a)$ painted darker), while
it is left blank otherwise.
 
(a) $a=1.71,N=300$. Every  128th step is plotted.   
(b) $a=1.73,N=300$. Every  256th step is plotted.
(c) $a=1.76,N=200$. Every  2048th step is plotted. 


\vspace*{.2in}
Fig.20: Lyapunov spectra of our model with $\epsilon = .5$, starting with
a random initial condition:  Comparison of quasistationary states with an 
attractor.  In the former, two sets of spectra in the transients states are
calculated after discarding 10000 initial transients, for
two different initial conditions.  They are overlayed, but agree within the
linewidth of the figure.  For the latter,
the data after 1000000 steps are adopted. The 
calculation is carried out through the products of Jacobi matrices over
32768 time steps. $N=50. a=1.88$
 
\vspace*{.2in}

Fig.21:  Switching process;
Amplitude-space plot of $x_n(i)$ with a shift of time steps.  200 sequential 
patterns $x_n(i)$ are displayed with time (per 128 time steps),
after discarding 40960 initial transients, starting from a random initial 
condition.  At the time steps and lattice point indicated by the arrows,
external input is applied to change the value of $x_n(i)$ at the 
corresponding $i$ and $n$ .
$\epsilon =.5$, and $N=64$
(a)$a=1.7$ (b)$a=1.75$.  

\vspace*{.2in}

Fig.22: Amplitude-space plot of $x_n(i)$ with a shift of time steps. 200 
sequential patterns $x_n(i)$ are plotted with time (per 1024 time steps),
after discarding 20480 initial transients, starting from a random initial 
condition.  
$\epsilon =.5$, and $N=100$. (a)$a=1.47$ (b)$a=1.52$.  
\vspace*{.2in}

Fig. 23: Spatial return maps with the corresponding space amplitude plots: 
$(x_n(1),x_n(2))$, $(x_n(13),x_n(14))$,  $(x_n(25),x_n(26))$, and
$(x_n(37),x_n(38))$ 
are plotted over the time steps $n=12800, 12801, \cdots ,64000$, 
while $x_n(i)$ is plotted with time per 256 steps.
$a=1.5, \epsilon =.5$, and $N=50$; (a)without any phase slip.
(b)with one phase slip.

\vspace*{.2in}

Fig. 24:  Spatial power spectra $S(k)$ obtained from
the Foureir transform of pattern $x_n (i)$.  Calculated from
the average over 100000 time streps after discarding  initial 10000 steps.
$\epsilon =.5$, and $N=2048$; (a)$a=1.5$ (b)$a=1.65$.

\vspace*{.2in}
Fig.25: Lyapunov spectra of our model with $\epsilon = .5$, starting with
a random initial condition.  Three examples of calculation are overlayed
starting from 3 randomly chosen initial conditions,
after discarding 10000 initial transients.
Calculated over 32768 time steps. $N=50$. $a=1.53$.
 
\vspace*{.2in}

Fig. 26: Co-moving mutual information flow for the logistic lattice (1),
obtained with the algorithm in the text. $I(m,t_c \times j)$ is plotted 
for $-10 \leq m \leq 10$ and $0 \leq j \leq 16 $ with coarsegraining time
$t_c=256$. 
The probability is calculated using 64 bins and
sampled over 5000x$t_c$ steps over the whole lattice.
$a=1.5, \epsilon=.5$ and $N=100$

\vspace*{.2in}
Fig. 27   Amplitude-space plot of $x_n(i)$ (for a moving kink)
with a shift of time steps.  200 sequential 
patterns $x_n(i)$ are depicted with time (per 1024 time steps),
after discarding 4096 initial transients, starting from a random initial 
condition.  At the time steps and lattice points indicated by the arrows,
the value of $x_n(i)$ at the corresponding $i$ and $n$ is shifted to 0 by an input.
$a=1.4$, $\epsilon =.5$, and $N=64$.


\addcontentsline{toc}{section}{References}
 
\begin{thebibliography}{999}
 
\bibitem{KK-PRL}
Rapid communication of the present studies is given in
K. Kaneko, Phys.Rev. Lett. 69 (1992) 905 

\bibitem{KK-PTP}
K. Kaneko, Prog. Theor. Phys. 72  (1984) 480, 74 (1985) 1033
 
\bibitem{KK-book}
K. Kaneko, Ph. D. Thesis {\sl Collapse of Tori and Genesis of
Chaos in Dissipative Systems}, 1983 (enlarged version is
published by World Sci. Pub., 1986)
 
\bibitem{KK-Physica}
K. Kaneko, Physica 23D  (1986) 436


\bibitem{KK-PD}
K. Kaneko, Physica 34D (1989) 1; 36D (1989) 60

\bibitem{JPCKK}
J. P. Crutchfield and K. Kaneko,``Phenomenology of Spatiotemporal Chaos", 
in {\sl Directions in Chaos} (World Scientific, 1987)
 
\bibitem{KK89d} 
K. Kaneko ``Simulating Physics with Coupled Map Lattices
------ Pattern Dynamics, Information Flow, and Thermodynamics of
Spatiotemporal Chaos" 
in {\sl Formation, Dynamics, and Statistics of Patterns}
(ed. K. Kawasaki, A. Onuki, and M. Suzuki, World. Sci. 1990)

\bibitem{Kapral}
I. Waller and R. Kapral, Phys. Rev. 30A  (1984) 2047;
R. Kapral, Phys. Rev. 31A  (1985) 3868
 
\bibitem{INT}
for type-I spatiotemporal intermittency, see [1],\cite{JPCKK}, and
H. Chat\'{e} and P. Manneville, Physica 32D (1988) 409.

for type-II intermittency, see J. D. Keeler and J. D. Farmer, Physica 23D 
(1986) 413; \cite{KK-PD}.  

See also P. Grassberger and T. Schreiber, Physica 50 D (1991) 177

\bibitem{TRS}
J.P. Crutchfield and K. Kaneko, Phys. Rev. Lett. 60 (1988) 2715 and in 
preparation: K. Kaneko, Phys. Lett. 149A (1990) 105

\bibitem{KK-THERMO}
K. Kaneko, Prog. Theor. Phys. Suppl. 99 (1989) 263


\bibitem{KK-D}
R.J. Deissler and K. Kaneko, Phys. Lett. 119A (1987) 397

 

\bibitem{CI}
K. Kaneko, Phys. Rev. Lett. 63 (1989) 219; Physica 41D (1990) 38;
K. Ikeda, K. Matsumoto, and K. Ohtsuka, Prog. Theor. Phys. Suppl. 99 (1989)
295;
I. Tsuda, in {\sl Neurocomputers and Attention}, (eds.  A.V. Holden and 
V. I. Kryukov, Manchester Univ. Press, 1990)

\bibitem{Fred}
F. Willeboordse, Phys. Rev. E (Brief Reports), in press
(1992); F. Willeboordse, PhD thesis, Tsukuba Univ., to be submitted

\bibitem{Freeman}
W. Freeman and C. A. Skarda, Brain Res. Rev. 10 (1985) 147;
W. Freeman, Brain Res. Rev. 11 (1986) 259

\bibitem{Shaw}
R. Shaw, Z. Naturforshung, 36a (1981) 80;
{\sl The Dripping Faucet as a Model Chaotic System},
(Aerial Press, Santa Cruz, 1984)

\bibitem{Ikeda}
K. Ikeda and K. Matsumoto, Phys Rev Lett. 62 (1989) 2265


\bibitem{kink}
This type of moving kink domains is first discussed in C. Bennett, G. Grinstein,
Y. He, C. Jayaprakash and D. Mukamel, Phys. Rev.  A 41 (1990) 1932.  
In the case of period-2 attractor, possible attributed phases are
$\pi$ or $-\pi$.  Thus it is impossible to distinguish between the
phase increase or decrease (for example  the phase changes with the
sequence of $\pi ,-\pi ,\pi$ can have neither phase advance or retardation.)

\bibitem{Kolo}
D.R. Ohlsen et al., Phys. Rev. Lett. 65 (1990) 1431;
P. Kolodner, Phys. Rev. A 42 (1990) 2475 ;
V. Croquette and H Williams Pjysica 37 D (1989) 295;
D. Bensimon et al., J. Fluid Mech. 217 (1990) 441

\bibitem{Kolo2}
P. Kolodner, J.A. Glazier, and H Williams, Phys. Rev. Lett. 65 (1990) 1579 
 
\bibitem{Croquette}
V. Croquette, J.M Flesselles and S. Jucquois, private communication
 
\bibitem{Yana}
Yanagita has recently observed the traveling wave in B\'{e}nard convection
in the CML model for convection given by T. Yanagita and K. Kaneko, 
"CML model for convection", submitted to Phys. Lett.  A.
 
\end{thebibliography}
\end{document}